\documentclass{article}
\hoffset= - .8in
\def\half{{\textstyle{1\over2}}}
\def\2{{1\over 2}}

\def\d{\partial}
\newcommand{\rf}[1]{(\ref{#1})}
\usepackage{amssymb}

\title{\begin{flushright}
{\normalsize ITEP/TH-59/05\\
FIAN/TD-16/05}
\end{flushright}
\vspace{10mm}
\bf{On First Order Formalism in String Theory}}
\author{Andrei S. Losev$^1$\footnote{losev@itep.ru}, Andrei Marshakov$^{2,1}$
\footnote{mars@lpi.ru, mars@itep.ru },
Anton M. Zeitlin$^3$\footnote{zam@math.ipme.ru, http://www.ipme.ru/zam.html}\\
$^1$ Institute of Theoretical and Experimental Physics,\\
B. Cheremushkinskaya 25,
Moscow 117259, Russia\\
$^2$Department of Theoretical Physics,
Lebedev Physics Institute,\\
Leninsky pr. 53,
Moscow 119991, Russia\\
$^3$St. Petersburg Department of Steklov Mathematical
Institute,\\
Fontanka, 27,
St. Petersburg, 191023, Russia}
\date{}

\begin{document}
\maketitle
\begin{abstract}
We consider the first order formalism in string theory, providing a
new off-shell description of the nontrivial backgrounds
around an "infinite metric". The OPE of the vertex operators, corresponding
to the background fields in some "twistor representation", and
conditions of conformal invariance results in the quadratic equation
for the background fields, which appears to be equivalent to the Einstein
equations with a Kalb-Ramond B-field and a dilaton. Using a new representation
for the Einstein equations with B-field and dilaton we find a new class of
solutions including
the plane waves for metric (graviton) and the B-field.
We discuss the properties of these background equations and main features of the
BRST operator in this approach.
\end{abstract}
\section{Introduction}
\hspace*{5mm}
String theory in non-trivial background is a very complicated problem.
In contrast to the flat-space case, where the perturbative
amplitudes can be computed by calculation
of the Gaussian integrals, generally one has to study a non-trivial sigma-model, which
is rarely equivalent to an exactly solvable
two-dimensional conformal field theory \cite{Polbook}. The connection between
the clear space-time sigma-model picture and axiomatically formulated two-dimensional
conformal field theory is often hidden, and sometimes is not even clear on fundamental
level -- as in the case of $AdS_5\times S^5$ \cite{AdS5} and pp-wave backgrounds
\cite{Metsaev} for ten-dimensional
superstring. One of the possible ideas (with recently renewed interest) is that
string theory of gravity can be more successful,
being considered in vicinity of a background, which is singular from conventional point
of view of classical gravity. For such backgrounds it is not even clear, whether a
traditional sigma-model formalism can help in finding a two-dimensional conformal
field theory, corresponding to their quantization.

String theory is usually formulated as a perturbative expansion of certain nonlinear
"string field theory" around some background
classical solution to its equations of motion. Even suppressing all string loops and in
the limit of vanishing string length $\alpha'\to 0$
it has to reproduce the highly nonlinear Einstein equations on the background fields,
containing all powers of perturbation, being expanded around the flat-space background.
Since perturbative expansion generally depends on the background, it seems reasonable
to start with studying some simple ones for this purpose. In this
paper we propose to start with a kind of "Gromov-Witten" background \cite{witten2},
with the
infinite target space metric and the $B$-field $G_{i\bar j}=\pm B_{i\bar j}$:
it turns out that theory drastically simplifies in this limit, and can be described
in terms of a conformal {\em first}-order system.

Hence, we are going to study the first-order formalism in string theory,
based on
the $D/2$-tensor power of the free $c=2$ conformal field theory, interacting
with reparameterization ghosts; during past years different two-dimensional first order
field theory models were extensively studied, see e.g. \cite{felder}-\cite{frenkel}.
We demonstrate, that this formalism describes the string theory around the
infinite-metric background (accomplished with the infinite B-field), which, being defined
precisely, necessarily requires the target-space complex structure.
One of the main advantages of this formalism is that the vertex operator
perturbations of the first-order action correspond generally to the {\em off-shell}
background fields in the "light sector" of the target-space theory. Due to absence
of the self-contractions between the co-ordinate fields themselves,
most anomalous dimensions vanish and there are no
higher-spin states, causing complications in conventional formulation
of the theory.

In order to get generic background, one has to perturb this theory by the set of all
marginal fields, and
the vertex operator perturbations of the first-order theory are adequately formulated
in terms of certain "twistor" variables $g^{i\bar j}$, $\mu_{i}^{\bar j}$ (together with
its complex conjugated $\mu_{\bar i}^{j}$) and $b_{i\bar j}$, which have clear
algebraic origin and
whose connection with physical background
fields $G_{\mu\nu}$, $B_{\mu\nu}$ and dilaton field $\Phi$ is rather nontrivial.
We study conditions for these fields to be marginal and exactly marginal
(absence of the $1/|z|^2$-terms in the the operator product expansions (OPE)
of the vertex operators) and derive
the field equations of motion. These constraints in the first-order formalism
appear to have more rich structure, than in conventional sigma-model approach and
we analyze the resulting equations of motion, in particular the bilinear equation to the
inverse metric $g^{i\bar j}$ (see equation \rf{cond1}) from the point of view
of target-space gravity and algebraic structure of the theory.
Even restricted to a very special class of perturbations of the form
$g^{i\bar{j}}p_i p_{\bar{j}}$ we obtain the set of linear and quadratic equations for the
background fields, whose solutions (together with conditions that the background fields are
primary) appear to be solutions of the full nonlinear system of Einstein equations for
the background physical fields\footnote{One could expect this generally, since cubic
terms in these equations should be of higher order in $\alpha'$.}.

Finally we are going to discuss briefly some non-perturbative aspects of possible
application
of the first-order formalism. In particular, we note that disappearance of the higher-spin
fields and conventional on-shell condition was observed recently in \cite{PR}
where the infinite metric of the AdS-like backgrounds was generated by thick stack of
D-branes.
In this paper we will not really go beyond the quadratic
approximation and only briefly speculate on possible BRST structure of the model.

\section{The first-order theory}

\hspace*{5mm}
Let us start with a two-dimensional conformal field theory (CFT) with the
first-order action \cite{witten2}:
\begin{eqnarray}\label{free}
S_0=\frac{1}{2\pi\alpha'}\int_\Sigma d^2 z (p_i\bar{\partial} X^{i}+
p_{\bar{i}}{\partial} X^{\bar{i}}),
\end{eqnarray}
where the momentum $p,\bar{p}$-fields are the $(1,0)$- and $(0,1)$-forms correspondingly,
while the co-ordinates $X,\bar{X}$ are scalars with the weights $(0,0)$, the volume
element is $d^2 z=i dz\wedge d\bar{z}$. Action \rf{free} corresponds
to the $D/2$-th tensor power of (holomorphic and anti-holomorphic) $c=2$ first-order
conformal field theory, where the multipliers are labeled according to some choice of
the target space complex structure, $i,{\bar i}=1,\ldots,D/2$.
The equations of motion following from the Lagrangian \rf{free} provide the
nontrivial operator product expansions (OPE):
\begin{eqnarray}
&&X^{i}(z_1)p_{j}(z_2)\sim\frac{\alpha'\delta^{i}_{j}}{z_1-z_2}+\ldots, \quad
X^{\bar{i}}(\bar{z}_1)p_{\bar{j}}(\bar{z}_2)
\sim\frac{\alpha'\delta^{\bar{i}}_{\bar{j}}}
{\bar{z}_1-\bar{z}_2}+\ldots
\end{eqnarray}
(it is convenient to keep here explicit $\alpha'$-dependence),
and there are no singular contractions between the $X$- and $p$-fields
themselves.

To study the theory with the action \rf{free}, let us, first, perturb it
by the following vertex operator:
\begin{eqnarray}\label{vertex}
&&V_g=\frac{1}{2\pi \alpha'} g^{i\bar{j}}(X,\bar{X})p_i p_{\bar{j}}
\end{eqnarray}
so that the full action becomes:
\begin{eqnarray}\label{fo}
S_{g}= \frac{1}{2\pi\alpha'}\int_\Sigma d^2 z (p_i\bar{\partial} X^{i}+
p_{\bar{i}}{\partial} X^{\bar{i}}-g^{i\bar{j}}p_i p_{\bar{j}}).
\end{eqnarray}

On classical level, solving equations of motion for $p,\bar{p}$,
one immediately finds that the action \rf{fo} is equivalent to:
\begin{eqnarray}
\label{phys}
\mathcal{S}=\frac{1}{2\pi\alpha'}\int_\Sigma d^2z
g_{i\bar{j}}\bar{\partial} X^i\partial X^{\bar{j}}
=\frac{1}{4\pi\alpha'}\int_\Sigma d^2 z
(G_{\mu\nu}+B_{\mu\nu})\partial X^{\mu}\bar{\partial}X^{\nu},
\end{eqnarray}
where $\mu,\nu$ run now over both holomorphic and antiholomorphic indices, while $G$,
and $B$ are the symmetric Riemann metric and antisymmetric Kalb-Ramond B-field
correspondingly. The physical fields
should obey the constraint $G_{i\bar{j}} = -B_{i\bar{j}}$, or
\begin{eqnarray}
G_{i\bar{k}}=g_{i\bar{k}}, \quad B_{i\bar{k}}=-g_{i\bar{k}}.
\end{eqnarray}

Note, that the operator \rf{vertex} contains the inverse metric $g^{i\bar j}$, written
in terms of the target-space holomorphic and anti-holomorphic co-ordinates, and, therefore,
is a perturbation of \rf{free} around the {\em infinite} metric background
(with the infinite Kalb-Ramond field).

However, in quantum case the integration measure should be taken into account.
For the first-order system \rf{fo} it is determined by the
holomorphic $D/2$-form $\Omega = \Omega(X)=dX^1\wedge\ldots\wedge dX^{D/2}$.
After integration over the $p$-fields
\begin{eqnarray}
\label{Adil}
\int[dp][d\bar{p}]e^{-S_{g}[X,\bar{X},p,\bar{p}]}
\sim
e^{-\mathcal{S}[X,\bar{X}]+\frac{1}{2\pi}\int_\Sigma d^2 z\sqrt{h} R\log\sqrt{g}}
\end{eqnarray}
we arrive at the standard sigma-model \rf{phys},
where the measure is determined with the help of
non-degenerate target-space metric.
The difference in two measures leads
to appearance of the dilaton $\frac{1}{2\pi}\int d^2 z\sqrt{h} R\log\sqrt{g}$
term in the action \rf{Adil},
related to determinant of the ultra-local operator.

Indeed,
integration in \rf{Adil} over the momenta
$p,{\bar p}$ naively leads to the (infinite) factor $ \prod_{z\in\Sigma} \det
g_{i\bar{j}}(X(z))$,
which plays a role of a factor, that turns the measure
determined by holomorphic form $\Omega$ into the measure determined by
non-degenerate metric $g$.
However, the "number of factors" in the infinite product
is the (infinite) number of one-forms, while the number
of factors needed to complete the measure on the $X$-fields equals to
the (infinite) number of functions (or zero-forms). It is well-known, that
difference between
these two infinite numbers is finite and equals to the arithmetic
genus $g-1$ of the world-sheet, or is proportional to the integral of the
scalar curvature along the surface. If we regularize this anomaly, say,
with the help of massive regulator fields, it becomes "locally distributed" along
the world-sheet, and this is a shortcut to understanding the dilaton term
(\ref{Adil}).

Another way to test the validity of (\ref{Adil})
is
to consider the target-space holomorphic
transformation $\delta X^i = \epsilon v^i(X)$,
$\delta p_i = -\epsilon p_j{\d v^j\over\d X^i}$.
The corresponding current $p_iv^i(X)$ obeys the anomaly equation
\begin{eqnarray}
\label{granom}
\bar\partial \left<p_i v^i(X)\right> = {1\over 2\pi}R\ \partial_i v^i(X) =
{1\over 2\pi}R\ \mathcal{L}_v\log\Omega
\end{eqnarray}
computed at some fixed point of the world-sheet.
That is in perfect agreement with (\ref{Adil}): one has to take into account
that $\det g$ is the ratio of two measures in the target-space,
determined by metric and by the holomorphic top form
$\Omega$ correspondingly.
The anomalous current \rf{granom} naturally suggests considering
the charges:
\begin{eqnarray}\label{charges}
n_v=
\frac{1}{2\pi i \alpha'}\oint_{S^1}dz\ v^i(X)p_i
\ \ \ \ \ \
r_{\omega}=
\frac{1}{2\pi i \alpha'}\oint_{S^1} dz\ \omega_i(X)\partial X^i
\end{eqnarray}
together with their complex conjugated $\bar n_v$ and $\bar r_\omega$,
generating the symmetries of the first-order action \rf{free}; their
properties are studied in Appendix A.

Now one can perturb the free action \rf{free} by all possible
operators of dimension (1,1), corresponding to more general deformation of metric,
$B$-field
as well as the deformation of the almost complex structure by the Beltrami differential
$\mu_{\bar{i}}^j$ and $\bar{\mu}^{\bar{j}}_i$. The full perturbed action reads
\begin{eqnarray}
\label{genpert}
&&S=\frac{1}{2\pi\alpha'}\int d^2 z (p_i\bar{\partial} X^{i}+
p_{\bar{i}}{\partial} X^{\bar{i}}\nonumber\\
&&-g^{i\bar{j}}p_i p_{\bar{j}}-\bar{\mu}^{\bar{j}}_i\partial X^i p_{\bar{j}}-
\mu_{\bar{i}}^j\bar{\partial} X^{\bar{i}} p_j - b_{i\bar{j}}\partial X^i
\bar{\partial} X^{\bar{j}}).
\end{eqnarray}
These background fields (to be called the twistor variables) can be
directly associated with the four independent terms in the expansion
(see formula \rf{bacalg} in Appendix A) of the tensor product of representation
spaces, corresponding
to action of the world-sheet symmetries \rf{charges} of the model.

Again, on classical level, solving equations of motion for $p,\bar{p}$, one finds
that this action is equivalent to the following sigma-model:
\begin{eqnarray}
\mathcal{S}=\frac{1}{2\pi\alpha'}\int d^2 z (g_{i\bar{j}}(\bar{\partial} X^i-\mu^{i}_{\bar{k}}
\bar{\partial} X^{\bar{k}})(\partial X^{\bar{j}}-\bar{\mu}_{k}^{\bar{j}}
\partial X^{k})-b_{i\bar{j}}\partial X^{i}\bar{\partial} X^{\bar{j}}),
\end{eqnarray}
which can be rewritten in the conventional form \rf{phys} with an extra 
dilaton term \cite{fts}
\begin{eqnarray}
\label{phys1}
\mathcal{S}=
\frac{1}{4\pi\alpha'}\int d^2 z
(G_{\mu\nu}+B_{\mu\nu})\partial X^{\mu}\bar{\partial}X^{\nu}+
\frac{1}{2\pi}\int d^2 z\sqrt{h}
R\Phi
\end{eqnarray}
with $G$,
$B$  and $\Phi$ now (compare to the previous section) defined as follows:
\begin{eqnarray}
\label{phytwi}
G_{s\bar{k}}&=&g_{\bar{i}j}
\bar{\mu}^{\bar{i}}_s\mu^{j}_{\bar{k}}+g_{s\bar{k}}-
b_{s\bar{k}}, \quad
B_{s\bar{k}}=g_{\bar{i}j}\bar{\mu}^{\bar{i}}_s\mu^{j}_{\bar{k}}-g_{s\bar{k}}-
b_{s\bar{k}}\nonumber\\
G_{si}&=&-g_{i\bar{j}}\bar{\mu}^{\bar{j}}_s-g_{s\bar{j}}\bar{\mu}^{\bar{j}}_i
, \quad
G_{\bar{s}\bar{i}}=-g_{\bar{s}j}\mu^{j}_{\bar{i}}-g_{\bar{i}j}\mu^{j}_{\bar{s}}
\nonumber\\
B_{si}&=&g_{s\bar{j}}\bar{\mu}^{\bar{j}}_i-g_{i\bar{j}}\bar{\mu}^{\bar{j}}_s,
\quad
B_{\bar{s}\bar{i}}=g_{\bar{i}j}\mu^{j}_{\bar{s}}-g_{\bar{s}j}\mu^{j}_{\bar{i}},
\nonumber\\
\Phi&=&\log \sqrt g.
\end{eqnarray}

\section{Main result}

\noindent
Let us now analyze the conformal invariance of the first-order theory, perturbed by
a single vertex operator $g^{i\bar{j}}p_i p_{\bar{j}}$ (\ref{vertex}).
The OPE of (\ref{vertex}) with the stress-energy tensor
$T=-(\alpha')^{-1}p_i\partial X^i$ (and its counterpart of
opposite chirality $\tilde{T}=-(\alpha')^{-1}p_{\bar{i}}\bar{\partial}X^{\bar{i}}$),
corresponding to the first-order system \rf{free} reads:
\begin{eqnarray}\label{primary}
-(\alpha')^{-1}p_i\partial X^i(z)\cdot(\alpha')^{-1}
g^{i\bar{j}}p_i p_{\bar{j}}(z') =
-{1\over (z-z')^3}\partial_ig^{i\bar{j}}p_{\bar{j}}(z') +\\
+{1\over\alpha'}\left({1\over (z-z')^2}g^{i\bar{j}}p_i p_{\bar{j}}(z')+
{1\over z-z'}\partial_{z'}g^{i\bar{j}}p_i p_{\bar{j}}(z')\right)
+ \ldots\nonumber
\end{eqnarray}
Two last terms in the r.h.s. of \rf{primary}, proportional to $(\alpha')^{-1}$,
are standard singular terms from
the OPE of the stress-tensor with primary field of unit dimension, so that, being integrated
over the world-sheet it becomes co-ordinate invariant, and they give no real constraints.
However the first
singular term in the r.h.s., proportional to $(\alpha')^{0}$, is
the action of the $L_1$-Virasoro operator, which deviates it
from the primary operator, unless
\begin{eqnarray}\label{gauge}
\partial_i g^{i\bar{j}}=0, \quad \partial_{\bar{j}}g^{i\bar{j}}=0
\end{eqnarray}
quite similarly to conventional "second-order" conformal field theory \cite{Polbook},
but arising here {\em before} any mass-shell condition, the same condition comes
from eliminating the contraction of $p_i$ and $X^j$ inside the operator (\ref{vertex})
$g^{i\bar{j}}p_i p_{\bar{j}}$ itself.
Moreover, in contrast to the second-order formalism, where transversality
justifies itself as a gauge artefact, being proportional to the two-dimensional
equations of motion and the total-derivative terms,
here the constraint (\ref{gauge}) appears as an independent requirement in the
target-space description of the theory.

Consider now the OPE of two vertex operators (\ref{vertex}) of the
general structure
$$
V(z_1)V(z_2)\sim\sum_{i,j}
\frac{a^{(i,j)}(z_2)}{(z_1-z_2)^{2-i}(\bar z_1-\bar z_2)^{2-j}}
$$
and calculate some important coefficients $a^{(i,j)}$.
For the vertex operators \rf{vertex} the most singular term in OPE
$a^{(0,0)}\propto(\alpha')^2 \d_k\d_{\bar l}g^{i\bar j}\d_i\d_{\bar j}g^{k\bar l}$
does not contribute in the leading order in $\alpha'$ and we will not discuss it
now.
The next is (the only at the level $(\alpha')^0$) logarithmic divergence,
coming from the double $p, X$ contractions, i.e.
\begin{eqnarray}
&&a^{(1,1)}=(2\pi^2)^{-1}(g^{i\bar{j}}\partial_{i}\partial_{\bar{j}}g^{k\bar{l}}-
\partial_{i}g^{k\bar{j}}\partial_{\bar j}g^{i\bar{l}})p_kp_{\bar{l}}+O(\alpha').
\end{eqnarray}
To make the theory conformally invariant it should vanish, i.e. the background
metric $g^{i\bar j}$ satisfies the "bilinear equation":
\begin{eqnarray}\label{cond1}
&&g^{i\bar{j}}\partial_{i}\partial_{\bar{j}}g^{k\bar{l}}-
\partial_{i}g^{k\bar{j}}\partial_{\bar j}g^{i\bar{l}}=0
\end{eqnarray}
In the case of general Hermitian metric\footnote{For the K\"ahler target-space
metric $g_{i\bar{j}}$ this condition leads to the vanishing Ricci tensor, and while gauge
condition (\ref{gauge}) is equivalent to the constant determinant, since
\begin{eqnarray}
0=\partial_ig^{i\bar{j}}g_{k\bar{j}}=-g^{i\bar{j}}\partial_ig_{k\bar{j}}=
-g^{i\bar{j}}\partial_kg_{i\bar{j}}=-\partial_k\log g \quad {\rm and} \quad c.c.
\end{eqnarray}}
we will show below that the conditions
of conformal invariance for the first order model (\ref{fo}) lead to the
background Einstein
equations with a dilaton, confirming the argument above.

This background equation is our new result, its algebraic properties are briefly
discussed in Appendix A; it is quadratic since the first-order theory corresponds
to expansion of the target-space theory around a singular background.

One can check (see Appendix B) that the quadratic system of equations \rf{cond1},
being supplied by the "gauge condition" (\ref{gauge}), is indeed equivalent to the
system of Einstein
equations with a Kalb-Ramond field and a dilaton \cite{fts}, \cite{eeq}:
 \begin{eqnarray}\label{einst1}
&&R_{\mu\nu}=-{1\over 4} H_{\mu\lambda\rho}H_{\nu}^{\lambda\rho}+2\nabla_{\mu}
\nabla_{\nu}\Phi,\\
&&\label{einst2}
\nabla_{\mu}H^{\mu\nu\rho}-2(\nabla_{\lambda}\Phi)H^{\lambda\nu\rho}=0,\\
&&\label{einst3}
4(\nabla_{\mu}\Phi)^2-4\nabla_{\mu}\nabla^{\mu}\Phi+
R+{1\over 12} H_{\mu\nu\rho}H^{\mu\nu\rho}=0.
\end{eqnarray}
where the change of variables from the "twistor variables" $g_{i\bar j}$ to
the physical metric, B-field and dilaton
$G, B, \Phi$ is given by the following expressions:
\begin{eqnarray}
G_{i\bar{k}}=g_{i\bar{k}},\quad B_{i\bar{k}}=-g_{i\bar{k}},\quad
\Phi=\log \sqrt{g}.
\end{eqnarray}
Note, that equivalence of the system
\rf{einst1}-\rf{einst3} to the equations \rf{cond1} and \rf{gauge}, coming directly
from OPE of \rf{vertex} in the first-order theory \rf{fo}, confirms the preliminary
conclusion of appearance of the dilaton from \rf{Adil}.

Let us stress again here, that the first-order theory corresponds to a singular-\-background
expansion of the Einstein equations (\ref{einst1})-(\ref{einst3}) and, therefore, in order
to make equivalence with the bilinear equation (\ref{cond1}) of the first-order theory one
has to use explicitly the gauge condition (\ref{gauge}), as required by conformal field
theory \rf{free}. In the common sigma-model approach, corresponding to expansion of the
action \rf{phys1} around a non-singular background, say
$G_{\mu\nu}=\eta_{\mu\nu}+h_{\mu\nu}$, the
"gauge" terms $\d_\mu h^{\mu\nu}=0$ can be eliminated by a prescription that the
terms proportional to the two-dimensional equations of motion and total derivatives are
cut off. However, in the singular background of first-order theory \rf{fo} there is no
linear approximation for the background field equations \rf{cond1}, i.e.
the "reference point" of expansion is singular from the point of view of the
target-space theory and this makes study of the symmetries in this point to be a delicate
issue.

We conjecture also that, as well as for the simplified model with the only perturbation
\rf{vertex}, the conformal invariance of the model in general background \rf{genpert}
at the order $(\alpha')^0$ is provided
by the Einstein equations with a $B$-field and a dilaton $\Phi=\log\sqrt{g}$.

\section{Special solutions}

An interesting question is to study the solutions of the system \rf{cond1}.
A particular class of solutions was discussed in \cite{HoTse}.
For example,
one can show that equations \rf{cond1} possess the following class of
solutions:
\begin{eqnarray}\label{sol}
g^{i\bar{j}}=\hat g^{i\bar{j}}(k_\mu X^\mu), 
\quad \hat g^{i\bar{j}}k_i  k_{\bar{j}}=0,\nonumber\\
k_i\left(\hat g^{i\bar{j}}\right)'=0 \rm{\quad and \quad c.c.},
\end{eqnarray}
where prime means the derivative of functions $\hat{g}^{i\bar{j}}(y)$
with respect to its argument $y=k_\mu X^\mu$. In the pure K\"ahler case one can write
a solution
\begin{eqnarray}
\hat g_{i\bar{j}}=\eta_{i\bar{j}}+k_i k_{\bar{j}}f(y),\ \ \ \ \ \
\eta^{i\bar{j}}k_i k_{\bar{j}}=0,
\end{eqnarray}
where $f(y)$ is any function of the scalar product $y=k_\mu X^\mu$.
Among these solutions one can find the plane waves, which are in physical
variables:
\begin{eqnarray}
G_{i\bar{j}}=\eta_{i\bar{j}} + e_{i\bar{j}}(A\cos(k_\mu X^\mu) +
B\sin(k_\mu X^\mu)),\nonumber\\
G_{i\bar{j}}=-B_{i\bar{j}}, \quad k_{\bar{l}}\eta^{i\bar{l}}e_{i\bar{j}}=0
\rm{\quad and \quad c.c.}
\end{eqnarray}
This means that the dilaton field, equal to $\log\sqrt{g}$, provides the plane
waves for the $G$- and $B$- fields.
One should also note that in our case $B$-field is pure imaginary (or
the two-form $g_{i\bar{j}}d z^i\wedge d z^{\bar{j}}$ is
anti-Hermitian). To make the $B$-field real, it is necessary to consider $z^i$ and
$z^{\bar{j}}$ not as complex conjugated variables, but as real ones
with $i, \bar{j}=1,\ldots,D/2$. Then the associated two-form becomes Hermitian,
but by obvious reasons the metric $g_{i\bar{j}}d z^i d z^{\bar{j}}$ acquires the
signature $(D/2,D/2)$.
An interesting feature of these solutions is that they do not
get additional $\alpha'$-corrections in the world-sheet perturbation
theory, since each loop diagram obviously contributes with the terms, vanishing due to
(\ref{sol}).

\section{Concluding remarks}

{\bf Remarks on D-branes near the AdS throat}.
One of the attractive special features of the proposed first-order formalism is
natural disappearance of the on-shell condition (as a linear equation on vertex
operator) together with the simultaneous disappearance of the higher-spin fields or Regge
descendants from the theory. Physically this phenomenon is a consequence of the infinite
metric limit, when the co-ordinate fields do not have contractions with themselves.
From the point of view of
two-dimensional conformal theory this kills the anomalous dimensions of the plane waves and
these  anomalous dimensions cannot compensate therefore
the dimensional polynomials of the derivatives of the co-ordinate
fields.

Similar phenomenon has been already observed in \cite{PR}, when D-brane is placed
in the vicinity
of the AdS throat. Clearly, near the throat the metric tends to infinity, and that explains the
observed in \cite{PR} effects in a rather similar way to how it happens the formalism
we have discussed in the paper.\\
{\bf Homotopic dreams}.
Our success in reproducing the solution to the Einstein equations \rf{einst1}-\rf{einst3}
in expansion around the singular first-order background motivates the study of all
perturbations, which should lead to the picture completely equivalent to the full set of
Einstein equations.
One can hope, that the Einstein equations in the language of string theory (in particular
of the proposed first-order theory) look like a kind of the Maurer-Cartan equation
\begin{eqnarray}
\label{mc}
Q\Psi + m_2(\Psi,\Psi) + m_3(\Psi,\Psi,\Psi) + \ldots = 0
\end{eqnarray}
where $Q$ is the BRST-operator in given background (see e.g. \rf{BRST}), $\Psi$ is
a (generalized!) vertex operator deforming the action, containing generally the
polyvertex fields, and $m_n(\Psi,\ldots,\Psi)$ are some operations in conformal theory,
corresponding to given background. The equations \rf{cond1}, \rf{gauge} we have derived,
correspond to
\begin{eqnarray}
\label{mceqmo}
&&Q\Psi =0,\nonumber\\
&&m_2(\Psi,\Psi) = 0
\end{eqnarray}
for the deformation \rf{vertex}. We expect that the conjectured set of equations \rf{mc}
would have a large symmetry group, promoting $\Psi \to \Psi + Q\epsilon$ to nonlinear
level, and that the operations $m_n(\Psi,\ldots,\Psi)$ would satisfy certain quadratic
equations like for homotopic structures.
We should also stress here that the conjectures higher operations are generally background
dependent and we hope that within the proposed first-order formalism they could appear
in the simplest possible form. We postpone the discussion of general deformation of the
BRST operator and structure of equation \rf{mc} for a separate publication.

\section*{Acknowledgements}
We are indebted to
V. Fock, A. Kapustin, N. Nekrasov, A. Niemi, B. Pioline, J. Schwarz, 
G. Semenoff, C. Vafa and, especially, to A. Tseytlin
for useful discussions. A.M.Z. is grateful to the Organizers of
Strings'05 Conference, Third Simons Workshop,
IAS and Perimeter Summer Schools and also to O. Aharony and
M. Berkooz (Weizmann Institute of Science) for kind hospitality, support and
possibility to have fruitful discussions.
The work of A.S.L. was supported
by RFBR grant 04-01-00637, Grant of Support for the Scientific Schools
1999.2003.2 and INTAS Grant INTAS-03-51-6346, the work of A.M. was partially supported by RFBR Grant 05-02-17451,
Grant of Support for the Scientific Schools 1578.2003.2, the NWO project 
"Geometric Aspects of Quantum Theory and Integrable 
Systems" 047.017.015, and by the Russian Science Support Foundation.
The work of A.M.Z. was supported by the Dynasty Foundation,
CRDF Grant No. RUMI-2622-ST-04 and RFBR grant 05-01-00922.

\section*{Appendix A. Algebraic structure of the first-order theory
\label{ss:alg}}

In this Appendix we consider the properties of the symmetries, generated by the
operators \rf{charges}.
The singularities coming from internal contractions \rf{granom} should be avoided
by the vanishing of divergences $\partial_i v^i=\partial_{\bar{i}} v^{\bar{i}}=0$.
Application of \rf{charges} to the vertex operator $V_g$ \rf{vertex} generates
the transformations
of fields, which is, to the order $\alpha'$:
\begin{eqnarray}\label{transf}
&&\delta_v g^{\bar{i}j}=-v^k\partial_k g^{\bar{i}j}-
\bar{v}^{\bar{k}}\partial_{\bar{k}}g^{\bar{i}j}+
\partial_{\bar{k}}\bar{v}^{\bar{i}}g^{\bar{k}j}+
\partial_{k}v^j g^{\bar{i}k}+O(\alpha')\\
&&\delta_{\omega}g^{\bar{i}j}=O(\alpha')\nonumber\\
&&\delta_{\omega} \mu^{j}_{\bar{i}}=\partial_{\bar{i}}\bar{\omega}_{\bar{k}}g^{\bar{k}j}-
\partial_{\bar{k}}\bar{\omega}_{\bar{i}}g^{\bar{k}j}+O(\alpha')\nonumber\\
&&\delta_{\omega}\bar{\mu}^{\bar{j}}_{i}=\partial_{i}\omega_{k}g^{k\bar{j}}-
\partial_{k}\omega_{i}g^{k\bar{j}}+O(\alpha'),\nonumber
\end{eqnarray}
i.e. $n_v$ generates
the holomorphic coordinate transformations, and $r_\omega$ is the generator of the
gauge symmetry $B\to B+D\omega+\bar{D}\bar{\omega}$, ($D$ and $\bar D$ are here
the {\em target-space} Doulbeaux operators),
which becomes clear, rewriting it for the $G$ and $B$ fields
(\ref{phytwi}):
\begin{eqnarray}
\delta_{\omega}B_{\mu\nu}=\partial_{\nu}\omega_{\mu}-\partial_{\mu}\omega_{\nu}+O(\alpha'),
\quad \delta_{\omega}G_{\mu\nu}=O(\alpha').
\end{eqnarray}
The algebra of the charges \rf{charges} is
\begin{eqnarray}
\label{algnr}
&&[n_{v_1},n_{v_2}]=n_{[v_2,v_1]}+\alpha'r_{\omega(v_1,v_2)}\\
&&[r_{\omega}, n_{v}]=r_{\mathcal{L}_v\omega},\quad [r_{\omega_1}, r_{\omega_2}]=0,\nonumber
\end{eqnarray}
where $\omega_n(v_1,v_2)=\frac{1}{2}(\partial_k v_2^l
\partial_n \partial_l v_1^k-\partial_n\partial_k v_1^l \partial_l v_2^k)$,
$\mathcal{L}_v\omega_k =\partial_i\omega_k v^i+\omega_i \partial_k v^i$
is a Lie derivative, and the same algebraic relations hold for the charges
of the opposite chirality $\bar n_v$, $\bar r_\omega$.
This is a deformation of the semidirect product
of the algebra of holomorphic coordinate transformations and the $B$-field gauge
transformations, where in the limit
$\alpha'\to 0$ the extension disappears.
One can also introduce a nondegenerate inner product, invariant under the adjoint
action $\mathcal{L}_v$:
\begin{eqnarray}\label{product}
(n_{v_1},n_{v_2})=0, \quad (r_{\omega_1},r_{\omega_2})=0,\ \ \ \
(n_{v},r_{\omega})=\int v^i(X)\omega_i(X)\Omega(X)
\end{eqnarray}
where $\Omega$ is a holomorphic volume form and the integral is taken along the
half-dimensional target-space cycle. It means that vector fields $v^i$ and one-forms
$\omega_i$  correspond to the dual representations $V$ and $V^\ast$
of the algebra \rf{algnr}
(already studied in \cite{schek}) and these representations provide a natural
algebraic structure of the background perturbation \rf{genpert}
\begin{eqnarray}\label{bacalg}
(V\oplus V^\ast)\otimes ({\bar V}\oplus {\bar V}^\ast)=
(V\otimes {\bar V}) \oplus (V\otimes{\bar V}^\ast) \oplus (V^\ast\otimes {\bar V})
\oplus (V^\ast\otimes {\bar V}^\ast)
\end{eqnarray}
where four terms in the r.h.s. literally correspond to the four background fields in
\rf{genpert}. Thus, we see, that generic perturbation of the first-order theory by
"light" fields has a natural algebraic origin.

The vertex operator $V$ or (\ref{vertex}) can be expanded
\begin{eqnarray}\label{expans}
V(X,p,\bar{X},\bar{p})=
\sum_{I}\mathcal{U}_{I}(X,p)\otimes
\bar{\mathcal{U}}_{I}(\bar{X},\bar{p})
\end{eqnarray}
in the (generally infinite; $I$ is some multi-index) bilinear combination of
the left- and right-chiral parts
$\mathcal{U}_{I}$ and $\bar{\mathcal{U}}_{I}$ with the
conformal weights $(1,0)$ and $(0,1)$ correspondingly.
In terms of \rf{expans} the transformation formulas (\ref{transf})
can be written in the form of adjoint action:
\begin{eqnarray}
&&\delta_vV=
\sum_{I}[n_v,\mathcal{U}_{I}]\otimes\bar{\mathcal{U}}_{I}+
\sum_{I}\mathcal{U}_{I}\otimes[\bar{n}_{\bar{v}},\bar{\mathcal{U}}_{I}]\nonumber\\
&&\delta_{\omega}V=
\sum_{I}[r_{\omega},\mathcal{U}_{I}]\otimes\bar{\mathcal{U}}_{I}+
\sum_{I}\mathcal{U}_{I}\otimes[\bar{r}_{\bar{\omega}},\bar{\mathcal{U}}_{I}]
\end{eqnarray}
Symmetries \rf{transf} are consistent with the conformal properties of the model
and, in the most compact way, this can be written as
\begin{eqnarray}
[Q,cn_v]=[Q,\tilde{c}\bar{n}_{\bar{v}}]=0, \quad
[Q,cr_{\omega}]=[Q,\tilde{c}\bar{r}_{\bar{\omega}}]=0
\end{eqnarray}
commutativity with the BRST operator \cite{brst} for the free first-order theory
(\ref{free})
\begin{eqnarray}
\label{BRST}
Q&=&\oint_{S^1}\mathcal{J},\quad \mathcal{J}=jdz-\tilde{j}d\bar{z},\\
j&=&cT+:bc\partial c:+\frac{3}{2}\partial^2 c, \quad \tilde{j}=\tilde{c}\tilde{T}
+:\tilde{b}\tilde{c}
\bar{\partial} \tilde{c}:+\frac{3}{2}\bar{\partial}^2 \tilde{c}\nonumber
\end{eqnarray}
where the $T=-(\alpha')^{-1}p_i\partial X^i$ and
$\tilde{T}=-(\alpha')^{-1}p_{\bar{i}}\bar{\partial}X^{\bar{i}}$ are correspondingly
holomorphic and antiholomorphic components of the energy-momentum tensor, and $b$, $c$
(and $\tilde{b}$, $\tilde{c}$) are reparameterization ghosts.

The OPE \rf{primary} can be also encoded into
the commutation relations of the BRST-operator \rf{BRST} with the fields,
which are:
\begin{eqnarray}
&&[Q,\phi_{h,\bar{h}}]=hc\partial\phi_{h,\bar{h}}+\bar{h}\tilde{c}\bar{\partial}\phi_{h,\bar{h}}+
\partial c\phi_{h,\bar{h}}+\bar{\partial}\tilde{c}\phi_{h,\bar{h}},\\
&&\lbrack Q,c\rbrack=c \partial c,\quad
\lbrack Q, b\rbrack=T+T^{gh},\ \ {\rm and}\ \ \ c.c.\nonumber
\end{eqnarray}
where $\phi_{h,\bar{h}}$ is a (primary) field with the
conformal weights $(h,\bar{h})$.
Denoting $c\tilde{c}V=\phi^{(0)}$, $V=\phi^{(2)}$, $cV-\tilde{c}V=\phi^{(1)}$, for
the vertex operator $V$ with conformal  weights $(h,\bar{h})=(1,1)$,
one can easily obtain the simple relations
$[Q,\phi^{(2)}]=d\phi^{(1)}$, $[Q, \phi^{(1)}]=d\phi^{(0)}$, $[Q, \phi^{(0)}]=0$
and $[Q,\int_M \phi^{(2)}]=\int_{\partial M}\phi^{(1)}$,
where $M$ is some two-dimensional manifold with a boundary.
Below we consider the
deformation of the BRST-operator for the nontrivial background in the first-order
theory and interpret the equations of motion for the background in terms of the
deformation theory for the BRST-operator \rf{BRST}.
We find that for generic perturbation this $L_1$-term should be supplemented
by the singular contributions from OPE of two vertex operators, giving rise to
(a linearized version of) some generalized Maurer-Cartan equation.

Equation (\ref{cond1}) can be also rewritten in the form:
\begin{eqnarray}
\label{condcV}
\lim_{\epsilon,\alpha'\to 0}\oint_{C_{\epsilon,z}}
\left(dz'\tilde{c}(\bar{z}')V(z')-d\bar{z'}c(z')V(z')\right)
c(z)\tilde{c}(\bar{z})V(z)=0,
\end{eqnarray}
where $C_{\epsilon,z}$ is a small contour around the point $z$, and this
form is used below for studying connection with the BRST operator. From
the point of view of sect.~\ref{ss:alg} the bilinear structure
and holomorphic properties of \rf{condcV} lead to appearance of a
{\em double}-commutator if one rewrites \rf{cond1} in the algebraic form.
Indeed, using \rf{expans} for the operator \rf{vertex}, one can write
for \rf{cond1}
\begin{eqnarray}\label{cond1alg}
&&g^{i\bar{j}}\partial_{i}\partial_{\bar{j}}g^{k\bar{l}}-
\partial_{i}g^{k\bar{j}}\partial_{\bar j}g^{i\bar{l}}=
\nonumber
\\
&&=\sum_{I,I'}\left((\mathcal{U}_{I}^{i}\partial_{i}\mathcal{U}_{I'}^{k})
(\mathcal{U}_{I}^{\bar j}\partial_{\bar j}\mathcal{U}_{I'}^{\bar l}) -
(\mathcal{U}_{I'}^{i}\partial_{i}\mathcal{U}_{I}^{k})
(\mathcal{U}_{I}^{\bar j}\partial_{\bar j}\mathcal{U}_{I'}^{\bar l})
\right),
\end{eqnarray}
where $\mathcal{U}_{I}^{i}=\mathcal{U}_{I}^{i}(X)$ and $\mathcal{U}_{I}^{\bar i}=
\mathcal{U}_{I}^{\bar i}(\bar X)$ are holomorphic and anti-holomorphic
"blocks" for the background metric field. Multiplying \rf{cond1alg} from the right by
$\partial_k\partial_{\bar l}$ one can rewrite \rf{cond1alg} as
\begin{eqnarray}\label{dcom}
&&\left(g^{i\bar{j}}\partial_{i}\partial_{\bar{j}}g^{k\bar{l}}-
\partial_{i}g^{k\bar{j}}\partial_{\bar j}g^{i\bar{l}}\right)\partial_k\partial_{\bar l}=
\nonumber
\\
&&=\sum_{I,I'}[{v}_{I}, v_{I'}]{\bar v}_I{\bar v}_{I'} =
{1\over 2}\sum_{I,I'}[{v}_{I}, v_{I'}][{\bar v}_I,{\bar v}_{I'}] = 0,
\end{eqnarray}
where we have introduced vector fields
$v_I=\mathcal{U}_{I}^{i}\partial_i$ and ${\bar v}_I=
\mathcal{U}_{I}^{\bar i}\partial_{\bar i}$.
For the r.h.s. of \rf{dcom} it is convenient to use the notation
\begin{eqnarray}\label{dcomm}
&&[[V,\tilde V]](X,p,\bar{X},\bar{p})=\sum_{I,J}
[\mathcal{U}_{I},\tilde\mathcal{U}_{J}](X,p)\otimes[\bar{\mathcal{U}}_{I},
\bar{\tilde{\mathcal{U}}}_{J}](\bar{X},\bar{p})
\end{eqnarray}
so that equation \rf{cond1} can be interpreted as vanishing of the double-commutator
\rf{dcom}, \rf{dcomm} in some algebra, naturally acting in the tensor product of the
holomorphic and antiholomorphic sectors of the first-order theory. We believe that this
is an algebraic structure naturally related with the theory of target-space gravity.

\section*{Appendix B. Relation between twistor and physical variables}
We use the formulas from Appendix $\Gamma$ of the book \cite{fock}:
\begin{eqnarray}\label{ricci}
R^{\mu\nu}=-\half  G^{\alpha\beta}\partial_{\alpha}\partial_{\beta}G^{\mu\nu}-
\Gamma^{\mu\nu}+\Gamma^{\mu,\alpha\beta}\Gamma^{\nu}_{\alpha\beta},
\end{eqnarray}
where
\begin{eqnarray}
&&\Gamma^{\mu\nu}=G^{\mu\rho}G^{\nu\sigma}\Gamma_{\rho\sigma},\quad
\Gamma_{\rho\sigma}=\half  (\partial_{\rho}\Gamma_{\sigma}+\partial_{\sigma}
\Gamma_{\rho})-\Gamma^{\nu}_{\rho\sigma}\Gamma_{\nu},\nonumber\\
&&\Gamma_{\nu}=G^{\alpha\beta}\partial_{\beta}G_{\alpha\nu}-
\half \partial_{\nu}\log (G).
\end{eqnarray}
Remember that in our case $\partial_{\mu}G^{\mu\rho}=0$, this
leads to the simple relation:  $\Gamma_{\nu}=-
\half \partial_{\nu}\log (G)$. Therefore $\Gamma_{\mu\nu}=-2\nabla_{\mu}
\nabla_{\nu}\Phi$, for the $\Phi=\log \sqrt{g}$, where $g$ is the determinant of
matrix $g_{i\bar{j}}$.
Now let us study the third term in (\ref{ricci}): first,
for the components of $ \Gamma^{\nu}_{\alpha\beta}$, one has:
\begin{eqnarray}\label{crist}
&&\Gamma^{i}_{rs}=\half  g^{i\bar{k}}(\partial_{r}g_{\bar{k}s}+
\partial_{s}g_{\bar{k}r}),  \nonumber\\
&&\Gamma^{i}_{r\bar{s}}=\half g^{i\bar{k}}(\partial_{\bar{s}}g_{r\bar{k}}
-\partial_{\bar{k}}
g_{r\bar{s}}) \rm{\quad and \quad c.c.}
\end{eqnarray}
while all other components vanish. Therefore, one finds that
$\Gamma_{\bar{i},r\bar{s}}=\half H_{\bar{s}\bar{i}r}$, hence the third term in
(\ref{ricci}) provides contribution of the $H^2$-type, with an additional
term in $\Gamma\Gamma$ for $\mu=\bar{i}$ and $\nu=j$:
\begin{eqnarray}
&&\Gamma^{\bar{i},kl}\Gamma^{j}_{kl}=-{1\over 4} (g^{k\bar{r}}\partial_{\bar{r}}
g^{l\bar{i}}+g^{l\bar{r}}\partial_{\bar{r}}
g^{k\bar{i}})g^{j\bar{p}}(\partial_{k}g_{\bar{p}l}+\partial_{l}g_{\bar{p}k})=
\nonumber\\
&&-{1\over 4} (g^{k\bar{r}}\partial_{\bar{r}}
g^{l\bar{i}}-g^{l\bar{r}}\partial_{\bar{r}}
g^{k\bar{i}})g^{j\bar{p}}(\partial_{k}g_{\bar{p}l}-\partial_{l}g_{\bar{p}k})-
g^{k\bar{r}}\partial_{\bar{r}}
g^{l\bar{i}}g^{j\bar{p}}\partial_{l}g_{\bar{p}k}=\nonumber\\
&&-{1\over 4}H^{\bar{i}kl}H^{j}_{kl}+\partial_{\bar{r}}g^{\bar{i}k}
\partial_{k}g^{\bar{r}j},
\end{eqnarray}
which, however, cancels with the first term in the r.h.s. of (\ref{ricci})
due to equation (\ref{cond1}).
Thus, unifying all the information we have got the relation (\ref{ricci}) can
be rewritten as:
\begin{eqnarray}
R^{\mu\nu}=-{1\over 4} H^{\mu\lambda\rho}H^{\nu}_{\lambda\rho}+2\nabla^{\mu}
\nabla^{\nu}\Phi.
\end{eqnarray}
Similarly, one can prove the following relation:
\begin{eqnarray}\label{aeinst}
4(\nabla_{\mu}\Phi)^2-2\nabla_{\mu}\nabla^{\mu}\Phi-
{1\over 6}H_{\mu\nu\rho}H^{\mu\nu\rho}=0.
\end{eqnarray}
Namely, let us start with $\partial_{\mu}\Phi=\half g^{\bar{i}k}\partial_{\mu}g_{\bar{i}k}$,
i.e.
\begin{eqnarray}\label{1}
2(\nabla_{\mu}\Phi)^2=g^{\bar{l}k}\partial_{i}g_{\bar{l}k}
g^{i\bar{j}}g^{\bar{s}r}\partial_{\bar{j}}g_{\bar{s}r}
\end{eqnarray}
and
\begin{eqnarray}\label{2}
-\nabla_{\mu}\nabla^{\mu}\Phi=-g^{\bar{i}j}\partial_{\bar{i}}
(g^{\bar{l}k}\partial_{j}g_{\bar{l}k})+g^{\bar{i}j}\Gamma^r_{\bar{i}j}
g^{\bar{l}k}\partial_{r}g_{\bar{l}k}+g^{\bar{i}j}
\Gamma^{\bar{r}}_{\bar{i}j}g^{\bar{l}k}\partial_{\bar{r}}g_{\bar{l}k}.
\end{eqnarray}
Using (\ref{crist}) we arrive at
\begin{eqnarray}\label{scrist}
g^{\bar{i}j}\Gamma^r_{\bar{i}j}=-\half g^{\bar{i}j}\partial_{\bar{l}}g_{\bar{i}j}
g^{\bar{l}r},\quad
g^{\bar{i}j}\Gamma^{\bar{r}}_{\bar{i}j}=
-\half g^{\bar{i}j}\partial_{l}g_{\bar{i}j}g^{\bar{r}l}.
\end{eqnarray}
The sum of (\ref{1}) and (\ref{2}) can be rewritten in the form:
\begin{eqnarray}
(\nabla_{\mu}\Phi)^2-\nabla_{\mu}\nabla^{\mu}\Phi=
-g^{\bar{i}j}\partial_{\bar{i}}
(g^{\bar{l}k}\partial_{j}g_{\bar{l}k}).
\end{eqnarray}
The $H^2$-term equals to:
\begin{eqnarray}
&&{1\over 6}H_{\mu\nu\rho}H^{\mu\nu\rho}=H_{i\bar{j}\bar{k}}H^{i\bar{j}\bar{k}}=
(-\partial_{\bar{k}}g_{i\bar{j}}+\partial_{\bar{j}}g_{i\bar{k}})
(-\partial_{s}g^{i\bar{j}}g^{s\bar{k}}+
\partial_{s}g^{i\bar{k}}g^{s\bar{j}})=\nonumber\\
&&=2g^{s\bar{k}}
\partial_{\bar{k}}g_{i\bar{j}}\partial_{s}g^{i\bar{j}}-2
\partial_{\bar{j}}g_{i\bar{k}}g^{s\bar{k}}\partial_{s}g^{i\bar{j}}.
\end{eqnarray}
One finds now, that (\ref{aeinst}) is satisfied due to (\ref{cond1}).
Combining (\ref{einst1}) and (\ref{aeinst}) one obtains (\ref{einst3}).

The third equation one can get by simple analysis of (\ref{cond1}). i.e.:
\begin{eqnarray}
\partial_{\bar{i}}(g^{\bar{i}j}\partial_{j}g^{\bar{k}l}-
g^{\bar{k}r}\partial_{r}g^{\bar{i}l})=0\rm{\quad and \quad c.c.}
\end{eqnarray}
leads to relation:
\begin{eqnarray}
\partial_{\bar{i}}H^{\bar{i}\bar{k}l}=0 \rm{\quad and \quad c.c.}
\end{eqnarray}
and identity
\begin{eqnarray}
\partial_{l}(g^{\bar{i}j}\partial_{j}g^{\bar{k}l}-
g^{\bar{k}r}\partial_{r}g^{\bar{i}l})=0\rm{\quad and \quad c.c.}
\end{eqnarray}
yields:
\begin{eqnarray}
\partial_{l}H^{l\bar{i}\bar{k}}=0 \rm{\quad and \quad c.c.}
\end{eqnarray}
These relations can be summarized as:
\begin{eqnarray}
\nabla_{\mu}H^{\mu\nu\rho}-2(\nabla_{\lambda}\Phi)H^{\lambda\nu\rho}=0.
\end{eqnarray}
Note here, that in the case of K\"ahler metric $g$ it is easy to show, that one
does not need
additional gauge constraint (\ref{gauge}) to prove the coincidence of equation
(\ref{cond1}) with the vacuum Einstein equation $R_{i\bar{j}}=0$.

\end{document}